# Ultrafast optical control of entanglement between two quantum dot spins

Danny Kim, Samuel G. Carter, Alex Greilich, Allan S. Bracker, Daniel Gammon
*Naval Research Laboratory*
*Washington DC 20375*

The interaction between two quantum bits enables entanglement, the two-particle correlations that are at the heart of quantum information science. In semiconductor quantum dots much work has focused on demonstrating single spin qubit control using optical techniques. However, optical control of entanglement of two spin qubits remains a major challenge for scaling from a single qubit to a full-fledged quantum information platform. Here, we combine advances in vertically-stacked quantum dots with ultrafast laser techniques to achieve optical control of the entangled state of two electron spins. Each electron is in a separate InAs quantum dot, and the spins interact through tunneling, where the tunneling rate determines how rapidly entangling operations can be performed. The two-qubit gate speeds achieved here are over an order of magnitude faster than in other systems. These results demonstrate the viability and advantages of optically controlled quantum dot spins for multi-qubit systems.

Semiconductor quantum dots (QDs) were among the first candidates proposed for solid-state qubits [1]. Self-assembled InAs QDs are a versatile physical platform, because they are epitaxially grown in a semiconductor wafer and can be fabricated into a monolithic architecture containing both electronic [2] and photonic [3] circuit elements. Individual QDs themselves can be organized into more complex "molecules" in one, two, and three dimensions. [4] With these engineering advantages, one can envision building an entire quantum network with the scalability and stability of a solid-state system.

The elementary optical excitation of a QD, the exciton, has a resonance frequency in the optical regime, giving QDs a great speed advantage over nuclear spins (radio frequency) or electron spins (microwaves). With the giant optical dipole of a semiconductor quantum dot, quantum operations can be performed at a terahertz rate or faster [5]. Coherent manipulations of pure exciton qubits were the first quantum gate demonstrations in the solid state [6, 7, 8, 9, 10, 11].

Unfortunately, the exciton lifetime is less than a nanosecond in a QD, so in the long run they are inadequate as qubits. This problem is solved by charging the QD with a single electron, and using the long-lived electron spin as the qubit. The electron spin lives for milliseconds [12] or longer and has a coherence time of microseconds [13, 14, 15, 16]. The exciton now acts as an auxiliary state for ultrafast conversion between optical coherence and spin coherence [17].

The opportunity created by this optically controlled spin paradigm has produced a burst of recent activity in single qubit operations with QDs, including single spin initialization [18, 19, 20], nondestructive readout [20, 21], and fast spin manipulation [22,23,24,25,26]. However, the power of quantum information originates from entanglement of multiple qubits, and so far all of these efforts are limited to a single isolated qubit. Creating and controlling entanglement requires a well-controlled spin interaction between QDs and quantum optical techniques that can address individual QDs within a pair. The interaction needs to be strong enough to maintain the speed advantage of optically controlled QDs, but not so strong that individual control of dots is impractical.

In this work, we go beyond one QD qubit for the first time, demonstrating the fabrication and optical control of a two-qubit system consisting of two electron spins, each on separate tunnel-coupled quantum dots. We establish concurrently the three important prerequisites for two-qubit entanglement control: (1) initialization of the two-electron spin state through optical pumping; (2) single qubit gates using short laser pulses; (3) and two-qubit gates using longer laser pulses. To achieve single qubit gates in a multiple qubit system, the key is a pulse so short that the two dots do not have time to interact while a single spin is optically rotated. For two-qubit gates we use pulses that are longer than the interaction time between the dots. Combining all these techniques, we obtain optical control of two quantum dot qubits on a time scale that is much faster than any other candidate for quantum computing [27,28,29].

## **TUNNEL-COUPLED QUANTUM DOTS**

Controlled entanglement of two spins requires an interaction between the spins, the strength of which determines how fast the two-qubit gate can be performed. A number of entanglement mechanisms have been proposed for optically driven QD spins, including coupling through photons and optical cavity modes [17, 30, 31], and through Coulomb and/or tunneling interactions between neighboring dots [32, 33, 34, 35, 36, 37]. Here we employ the kinetic exchange interaction based on coherent tunneling. Tunneling provides the largest interaction rate and because of recent advances in the growth and spectroscopy of vertically stacked QDs [38, 39,40,41] can be precisely controlled through a combination of quantum size engineering and applied electric field bias.

For this study two vertically stacked self-assembled InAs quantum dots (QDs) were grown with a thin tunnel barrier of GaAs/AlGaAs such that the electrons can coherently tunnel between the dots. We grew the two QDs with different thicknesses so that they have different optical transition energies. As a result each is optically addressable with a resonant laser frequency. The QDs are incorporated into a Schottky diode so that by adjusting the voltage bias each QD is charged with a single electron as shown in Fig. 1.

The calculated Hund-Mulliken model of the 2-$e$ energy level system [42] is shown in Fig. 2. Using optical transmission spectroscopy the 2$e$ charge state was found to be stable for a bias range of 100 mV centered at about 0.46 V. Beyond this range 1 and 3 electrons are more stable in the quantum dot pair. Within this 2$e$ stability range the lowest energy configuration consists of one electron residing in each dot. Because of the Pauli Exclusion Principle, the two electrons can only tunnel between QDs if they have an antisymmetric spin state. As a result the spin states are split into degenerate triplets ($T_0 = |\uparrow\downarrow\rangle + |\downarrow\uparrow\rangle$, $T_- = |\downarrow\downarrow\rangle$, $T_+ = |\uparrow\uparrow\rangle$) and a lower energy singlet ($S = |\uparrow\downarrow\rangle - |\downarrow\uparrow\rangle$) with a kinetic exchange splitting $\Delta_{ee}$ that is a result of tunneling between the dots [43]. The magnitude of the exchange energy determines the interaction strength between dots and ultimately how fast the entanglement can be controlled. We found that an exchange rate of 30 GHz enabled both ultrafast one- and two-qubit gates, as we will show.

The spin state is optically controlled through an exciton state ($X^{2-}$), in which an electron-hole pair is created in the top dot. Excitons in the smaller bottom dot are at a significantly higher energy. Tunneling between the two dots is rendered negligible in the excited state due to additional Coulomb interaction energies [42,44]. We start at zero magnetic field in which case the exciton state is four-fold degenerate, with the states labeled by their total spin projection $m_z$ in Fig. 2. The optical selection rules show that both $S$ and $T_0$ couple to the same optically excited state, forming a Λ-level diagram, as required for optical spin control.

## TWO-QUBIT INITIALIZATION: OPTICAL SPIN PUMPING

Optical spin initialization of the two-electron (2$e$) system is demonstrated by one-laser transmission spectroscopy. The two spectral lines in Fig. 3a (and lower spectra in Fig. 3b) arise from excitation of the 2$e$ singlet and triplet states to the exciton. The splitting between the lines is the kinetic exchange energy ($\Delta_{ee}$~120 μeV=29 GHz). Figure 3a shows bleaching of the singlet and triplet absorption lines in the center of the 2$e$ charge stability plateau, in the bias range between 0.45-0.48V. This is a well-known signature of optical spin pumping in a Λ-level diagram [18,19,20], where the spin is excited out of one spin state and becomes shelved in the other after a few recombination cycles. The pumping is efficient (at least 95% fidelity) as long as the optical cycle is faster than relaxation between the *S* and *T* ground states. On the edges of the stability plateau, spin-flip co-tunneling to the doping layer results in efficient ground-state relaxation, and no net pumping is possible. The spin pumping can be defeated by repumping on the opposite branch using a second laser as shown in the upper two linescans in Fig. 3b.

Interestingly, the selection rules of Fig. 2 (inset) indicate that two of the triplet sublevels ($T_+$ and $T_-$) are not part of a Λ-diagram and therefore should not pump at zero magnetic field. Nevertheless, the triplet absorption signal is almost completely suppressed through optical pumping. This implies that there is weak mixing of the triplets through the hyperfine interaction and/or heavy hole-light hole mixing in the exciton states. These interactions turn on additional optical transitions and allow full initialization into the singlet state. While spin initialization through optical pumping may seem routine, in this case incoherent pumping is in fact generating a particular *entangled* spin state, $S = |\uparrow\downarrow\rangle - |\downarrow\uparrow\rangle$. This entanglement is possible due to the exchange interaction. Using coherent optical techniques, this singlet state can be transformed into a variety of other entangled states.

**ENTANGLEMENT CONTROL USING PAIRS OF SINGLE QUBIT GATES**

For single qubit gates we use short, circularly polarized pulses (13 ps) with a bandwidth (146 μeV) larger than the exchange splitting. Because the pulses are shorter than the exchange interaction time and because the exciton is localized in the top dot, the pulse acts only on the spin in this dot. The pump laser pulses are detuned below the triplet transition by ~270 μeV, allowing a single spin rotation with only virtual occupation of the $X^{2-}$ exciton state [22,24,26]. This single qubit gate induces a relative phase between the up and down spin states in just the top dot – a rotation of this spin about the optical axis. As shown previously for a single electron spin, a circularly polarized π rotation takes $|\downarrow\rangle - |\uparrow\rangle$ into $|\downarrow\rangle - e^{i\pi}|\uparrow\rangle = |\downarrow\rangle + |\uparrow\rangle$, [22,24,25,26,45]. Similarly, in this two-electron case, where the pulse only acts on the top dot, the π-pulse rotates $S = |\uparrow\downarrow\rangle - |\downarrow\uparrow\rangle$ into $T_0 = |\uparrow\downarrow\rangle + |\downarrow\uparrow\rangle$.

As we now show, by applying two such single spin rotations separated by a delay time during which the 2$e$ spin state precesses, we can control the 2$e$ state. Initialization into the singlet state again is achieved with the $cw$ probe laser tuned to the triplet transition, which also acts as the measurement. At zero magnetic field, we operate in the $S$ and $T_0$ subspace, which can be visualized as a Bloch sphere as in Fig. 4a. The first pulse rotates the 2$e$ state from $S$ to a superposition with $T_0$, with a rotation angle that depends on the laser power. For example, an effective π/2- pulse rotates the Bloch vector to the equator as shown in Fig. 4a. Between pulses the state precesses around the Bloch sphere at the frequency of the exchange splitting ($\Delta_{ee}$). The second pulse drives the state either up or down depending on the phase of the superposition state. The projection onto $T_0$ determines the population and the corresponding signal of the $cw$ probe.

Figure 4b (upper curves) displays the oscillations in the signal as a function of pulse delay, called Ramsey fringes. The frequency of these oscillations (30 GHz) corresponds to the exchange energy $\Delta_{ee}$, which is a function of bias voltage. The observed frequencies as a function of bias are in excellent agreement with the measured singlet-triplet splitting (red dashed curve Fig. 4c). The oscillations have a decay time of 400-700 ps, although this can be substantially increased. The major damping source is found to be fluctuations in the bias voltage that produce fluctuations in the exchange energy. This is seen in Fig. 4c where the bias dependence of the decay has been fit to the derivative of the exchange energy ($\delta\Delta_{ee}/\delta V$) multiplied by the voltage fluctuation amplitude of 3.8 mV, taken as the fitting parameter. This contribution can be removed by designing the molecule such that $\delta\Delta_{ee}/\delta V = 0$ within the stability range of the 2$e$ configuration. Additional dephasing contributions due to the *cw* initialization/measurement laser that is persistently on can be eliminated by turning this laser off during the precession in between pulses [16]. Hyperfine induced coupling to the $T_+/T_-$ states is expected to lead to dephasing on a longer timescale, although a number of recent results on single QDs suggest that even this effect can be substantially reduced [14, 15, 13].

The amplitude of the oscillations as a function of pulse intensity is plotted in Figure 4d. The maximum amplitude occurs for two $\pi/2$-pulses, but decreases to nearly zero for two $\pi$-pulses, which should drive the system back to the initial *S* state, regardless of pulse delay. These oscillations as a function of laser power correspond to spin-Rabi oscillations, in which the Bloch vector can be driven all the way around the Bloch sphere. The amplitude dependence on the pulse intensity is not linear as expected from the adiabatic approximation but sublinear ($P^{0.7}$) as observed in single dots [24]. The combination of laser pulse area and delay time permits the controlled generation of any entangled superposition state within this part of the 2$e$ spin phase space, i.e. $\alpha|\uparrow\downarrow\rangle + \beta|\downarrow\uparrow\rangle$.

Optical control of the full 2$e$ spin state is possible using a transverse magnetic field that allows Raman transitions to the $T_+$ and $T_-$ states [46]. Figure 4b (lower curves) displays Ramsey fringes in a small transverse magnetic field of 0.2 T, which now show frequency components associated with $T_+$ and $T_-$. In a transverse magnetic field the short pulse now rotates *S* toward a superposition of $T_+$ and $T_-$, $|\uparrow\uparrow\rangle - |\downarrow\downarrow\rangle$, which precesses at twice the single electron Zeeman frequency. Calculated Ramsey fringes in this magnetic field are displayed in Fig. 4b and reproduce the experimental result very well. In this model, the populations of the $T_0$ and $T_\pm$ states after both pulses are $\frac{1}{2}W^2[1-\cos(\omega_+ - \omega_-)\Delta t]$ and $W(1-W)[1+\cos\omega_\pm \Delta t]$, respectively. The frequencies $\omega_\pm$ represent the $S$-$T_\pm$ splittings, $W$ is related to the single pulse area $\Theta$ through $W = \sin^2(\Theta/2)$, and $\Delta t$ is the delay between pulses. These results demonstrate the controlled creation of a superposition of *all four* 2e basis states, i.e. $\alpha|\uparrow\downarrow\rangle + \beta|\downarrow\uparrow\rangle + \gamma|\uparrow\uparrow\rangle + \delta|\downarrow\downarrow\rangle$.

## **TWO-QUBIT PHASE GATE**

It may seem surprising in these Ramsey fringe experiments that we can control the entangled state using only single qubit rotations. In reality the exchange interaction that causes precession on the Bloch sphere acts as a two qubit gate, with the timing between pulses effectively controlling the interaction time. We can also control this phase evolution with longer, narrowband pulses that act as two qubit phase gates. In analogy with pulse rotations developed for single QDs [45], a narrowband pulse near-resonant with an energy eigenstate that drives the system up to an excited state and then back down (a $2\pi$ pulse) induces a phase change. For example, an optical $2\pi$ pulse driving the singlet state gives $|\uparrow\downarrow\rangle - |\downarrow\uparrow\rangle \Rightarrow e^{i\phi}(|\uparrow\downarrow\rangle - |\downarrow\uparrow\rangle)$, where $\phi$ depends on pulse detuning [25, 26, 45]. When this gate with $\phi = \pi/2$ (often called square-root-of-swap) acts on a product state, $|\uparrow\downarrow\rangle$, the result is an entangled state, $|\uparrow\downarrow\rangle - i|\downarrow\uparrow\rangle$.

To obtain this two qubit phase gate, we first demonstrate optical Rabi oscillations (Fig. 5b). A pulsed laser with a relatively narrow bandwidth of 12 μeV (~150 ps pulsewidth), much less than the singlet-triplet splitting, is used to coherently drive the singlet transition. The *cw* probe is tuned to the triplet transition. As the pulse intensity increases, the triplet transmission signal shows Rabi oscillations that are periodic with the field amplitude (square root of the average laser power). The absorption signal does not return to zero due to recombination during these long pulses, as reproduced by a simulation using optical Bloch equations in a 3-level model (red curve in Fig. 5b).

The two-qubit phase gate is demonstrated by placing a $2\pi$ pulse, in this case near-resonant with the triplet transition, after the first short pulse of a Ramsey fringe experiment (see Fig. 5c). Pulse 1 rotates the Bloch vector up near the equator, where it precesses, and the longer control pulse changes the phase. Pulse 2 is used to measure these dynamics as a function of time by rotating the vector up or down, depending on its phase. In Fig. 5d, the oscillating signal goes nearly to zero during the control pulse, as the triplet population is excited to the exciton. As the exciton is driven back down to the triplet, the oscillations return with a phase change of 145°. This phase can be varied between -180° and 180° by changing the pulse detuning.

## **OUTLOOK**

The results presented in this article provide all of the ingredients for universal control of this two qubit system: initialization, single qubit gates, and an entangling two qubit gate. In this demonstration we have used QDs in which the coupling is in the ground state, and thus the spins are always interacting. Nevertheless, we can manipulate the spins independently by using ultrafast pulses that act faster than the interaction and because the optically excited states are localized. In analogy with NMR techniques, we should also be able to cancel the effect of the interaction at a given time using "refocusing" pulses if necessary [47]. Another approach would be to invert the system, *i.e.*, decouple the spins in the ground state and provide coupling in the exciton state [20, 41, 43]. The ultrafast optical techniques demonstrated here would be applicable as well if the interaction were only in the excited state.

Finally, we emphasize that one of the major advantages of this system is its natural coupling to photons, which can act as flying qubits for quantum communication. For example, the spin entanglement in the dots could be transferred to entanglement of emitted photons. Periodic excitation of the system should generate a two-dimensional cluster of entangled photons for measurement-based quantum computing [48]. Perhaps most exciting is the potential for further scaling using a quantum network, an approach to quantum computing or communication in which a set of nodes are entangled through quantum channels [30, 49]. These locally entangled spin qubits are ideal as a node in a distributed quantum network connected through photons [50]. For error suppression it is necessary to have within each node at least two qubits in which fast entanglement control is possible, as demonstrated here [50].

Acknowledgements
This work was supported by NSA/ARO, ARO MURI, DARPA, and ONR.

Figure Captions

Fig. 1 Diode structure and level diagram. Schematic diagram of device structure showing two electrons deterministically charged into separate dots. Lasers excite an electron-hole pair into the top dot to initialize or manipulate the spins.

Fig 2. Hund-Mulliken model of the 2e system showing possible orbital configurations of the ground and excited states, and their respective anticrossings. The red arrow shows approximately the bias in which these experiments were conducted. (Inset) Corresponding energy level diagram for spin substates showing nominal selection rules at zero magnetic field.

Fig 3. Optical spin pumping of the two electron state. (**a**) Intensity plot of the single laser transmission spectrum as a function of bias. The zero in the energy scale is at 1.3 eV. Dashed vertical lines indicate the edges of the two-electron stability plateau. (Inset) Lambda diagram with the singlet and triplet coupling to the same $X^{2-}$ state. (**b**) The lowest linescans are single laser spectra at two bias values: one in the co-tunneling regime that shows triplet and singlet peaks (0.51V) and one in the pumping regime where the peaks are gone (0.46V). The upper traces show that the transmission signals at 0.46V are restored by tuning the frequency of a repumping laser to the opposite arm of the Λ system.

Fig. 4. Coherent control of the 2e spin state using two 13 ps pulses with a variable time delay. (**a**) Bloch sphere representation of spin control within the S-$T_0$ subspace. The spin state is initialized to the south pole ($S = |\uparrow\downarrow\rangle - |\downarrow\uparrow\rangle$). The first pulse coherently drives the Rabi vector up after which it precesses around the sphere. The second pulse either drives the pulse up or down depending on the time separation between pulses and thereby leads to oscillations in the measurement. (**b**) Upper curve: the resulting Ramsey interference oscillations in the $T_0$ transmission signal as a function of the pulse separation time, with zero magnetic field. Lower curve: Ramsey interference obtained at 0.2T showing two frequency components. The fit is overlaid with S-$T_0$ splitting ~30GHz) and the slower $T_+/T_-$ splitting (2.2 GHz). The latter frequency is at twice the single electron Zeeman splitting ($g_e = 0.4$). (**c**) Measured Ramsey fringe frequency and decay time as a function of bias voltage at 0T. The dashed line for the fringe frequency is the measured singlet-triplet splitting ($\Delta_{ee}$) from Fig. 3a. The dashed line for the decay time is a calculation of the expected dephasing time due to voltage fluctuations of $\Delta_{ee}$ as discussed in the text. (**d**) Fringe amplitude vs. pulse intensity showing Rabi oscillations of the spin state.

Fig.5. Two qubit phase gate. (**a**) Lambda diagram showing the cw initialization/measurement laser and control pulse. (**b**) With a pulsed pump laser (~150 ps pulse width), optical Rabi oscillations are observed whose period scales with the square root of the pump intensity. (**c**) Pulse sequence for two-qubit phase gate experiment. (**d**) Ramsey interference oscillations with and without a two-qubit control pulse near-resonant with the triplet.

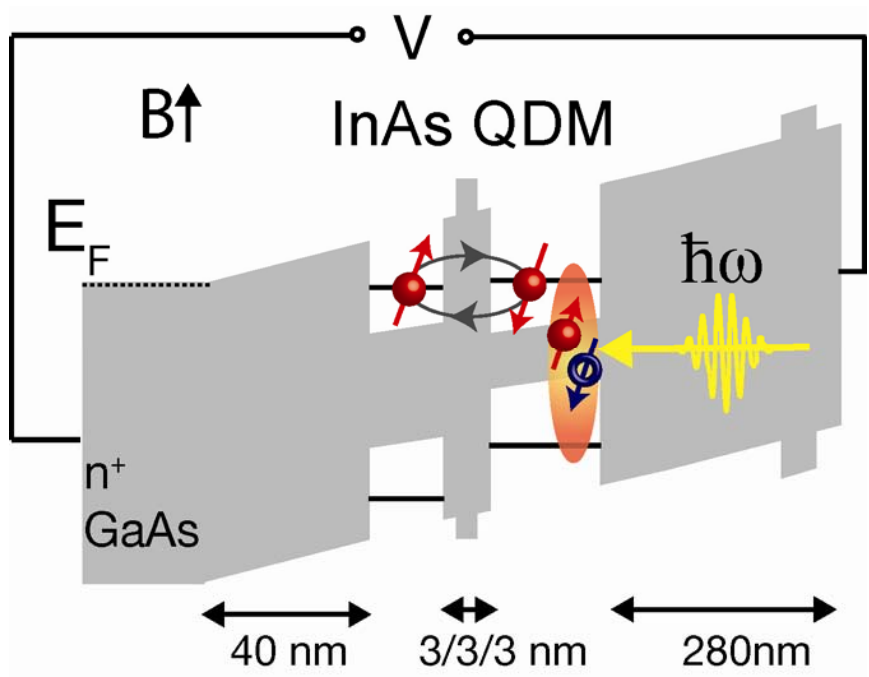

Fig 1.  D. Kim et al.

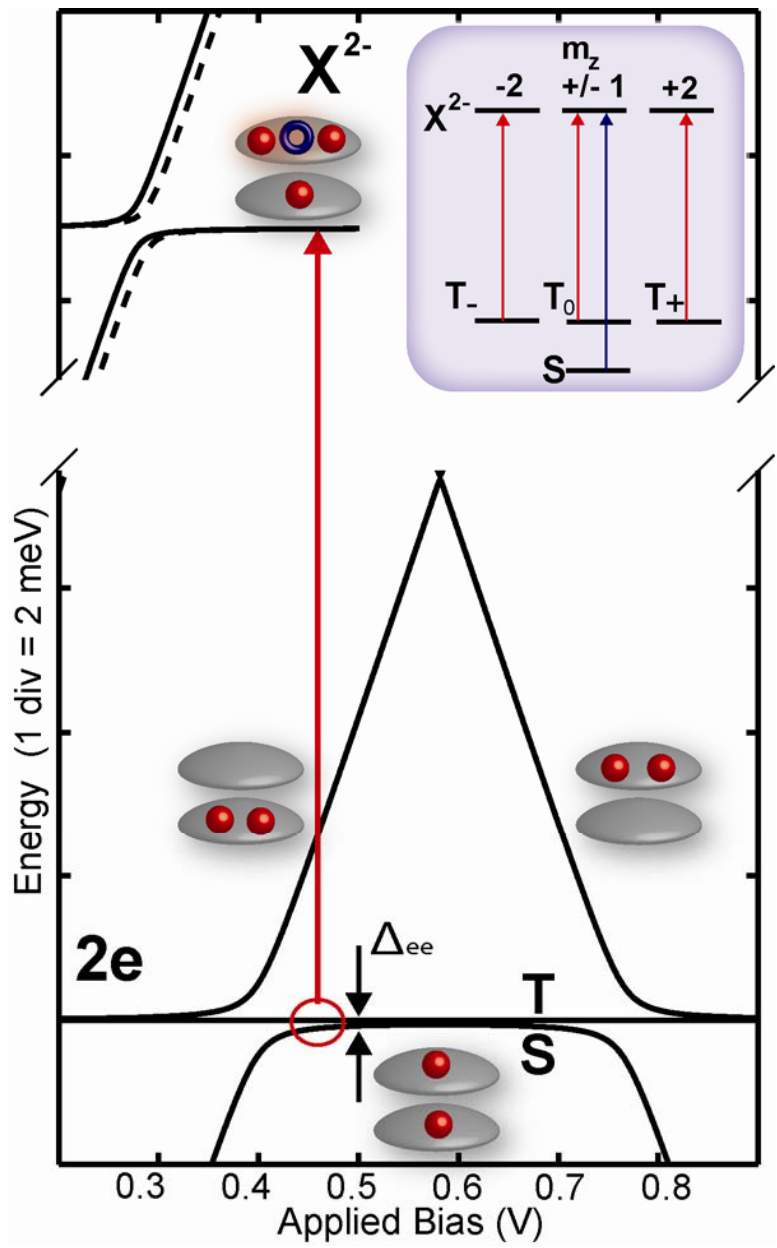

Fig 2. D. Kim et al.

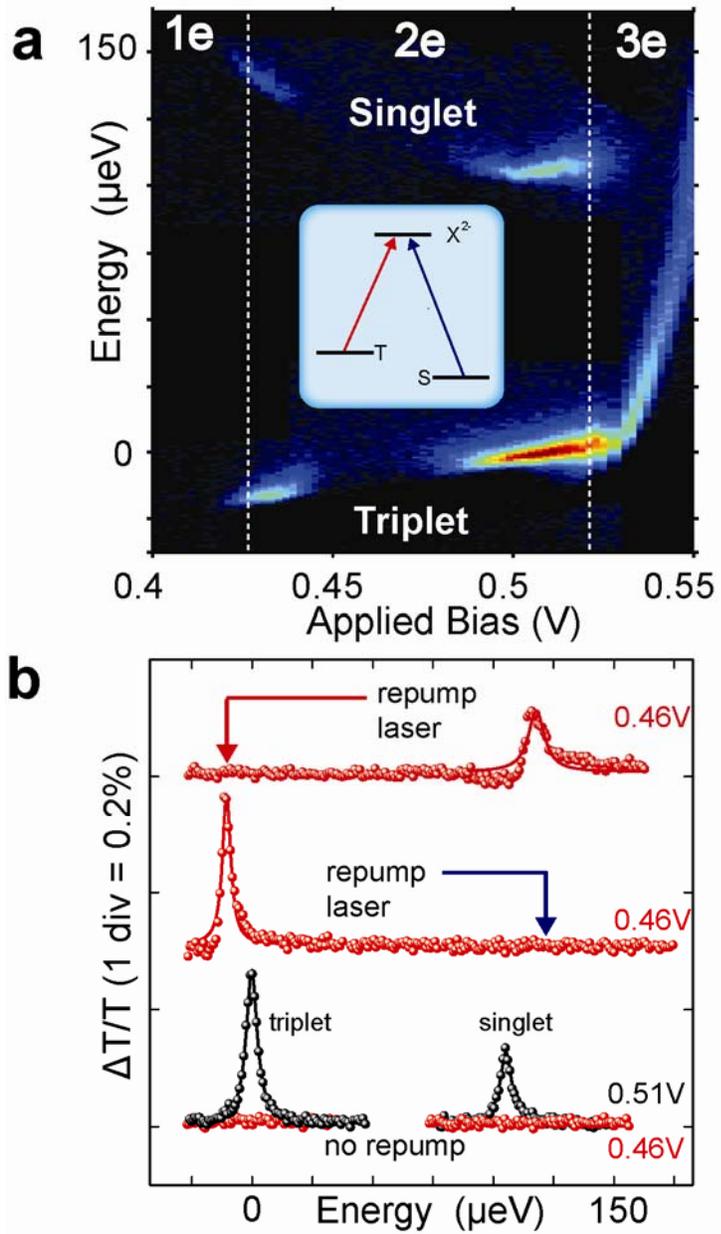

Fig 3. D. Kim et al.

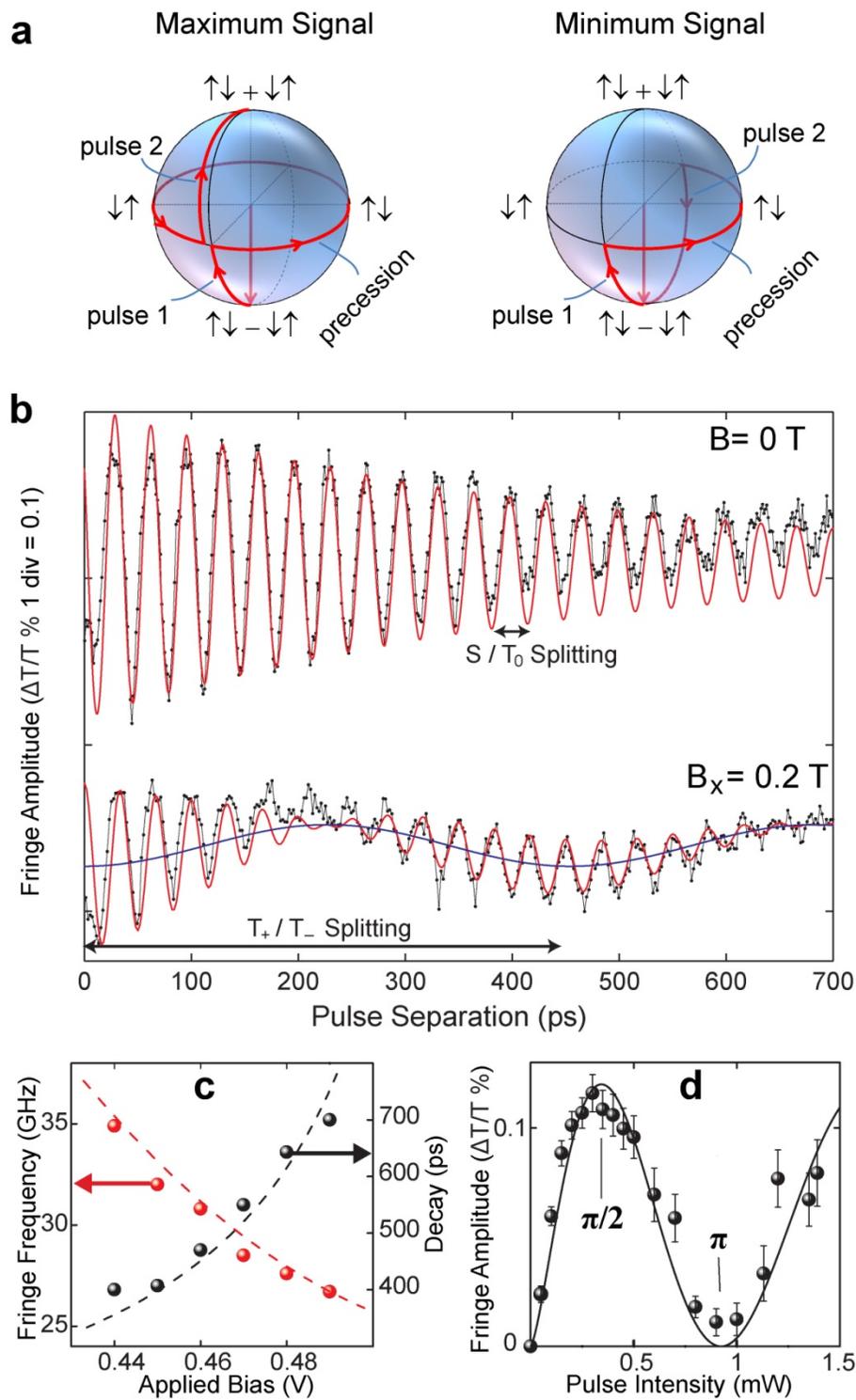

Fig 4  D. Kim et al.

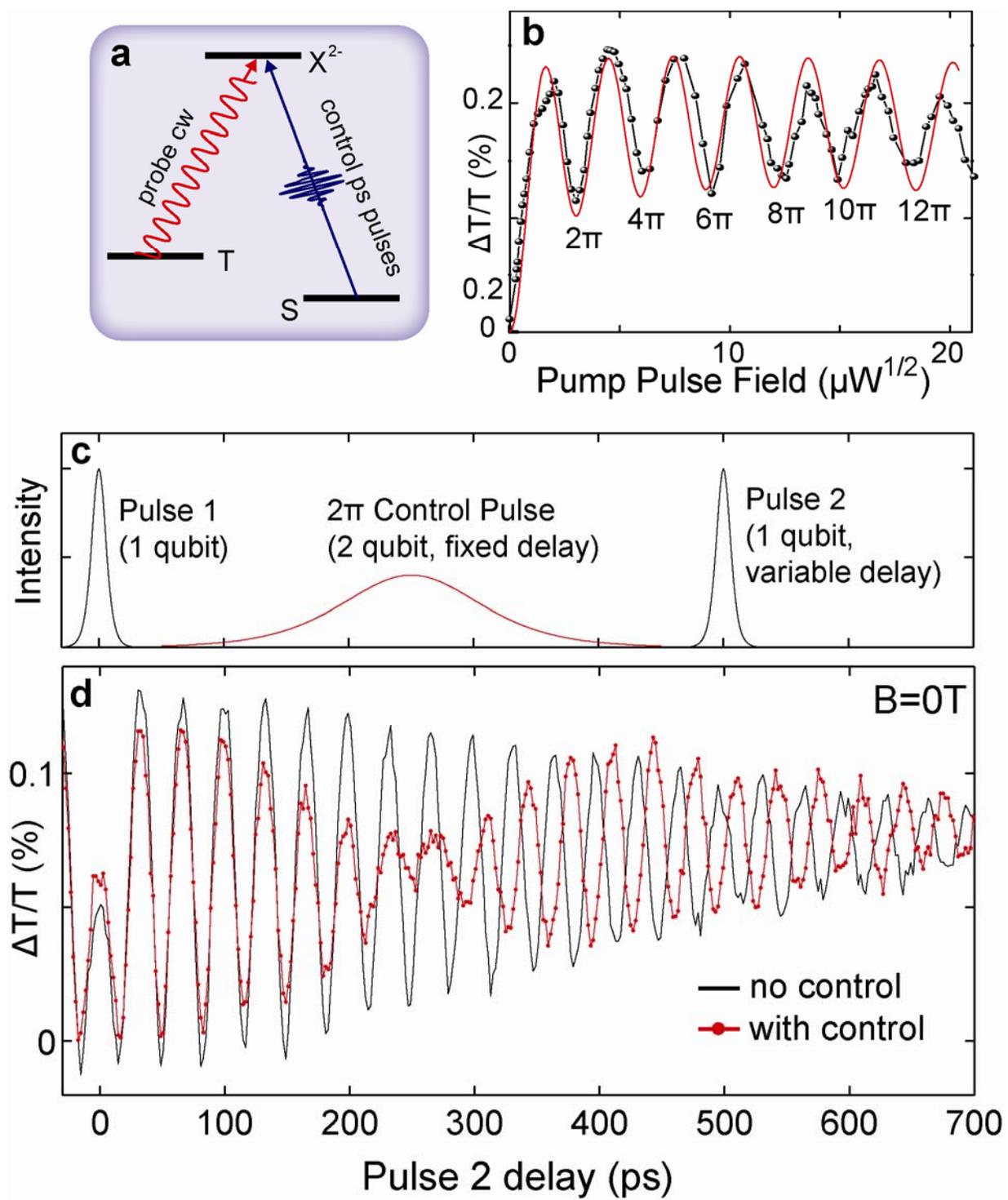

Fig 5   D. Kim et al.

**Methods Summary**

**Sample structure**

Two vertically stacked InAs dots were grown by molecular beam epitaxy on an n+ doped GaAs substrate. These pairs of QDs were grown with rotations and a chip with a low density of ~5 μm$^{-1}$ could be selected for single molecule studies. The sample consists of an n-type Si-doped buffer layer, a 40nm GaAs spacer barrier, the two InAs dot layers separated by a 9 nm GaAs/AlGaAs/GaAs barrier. A 280 nm GaAs capping layer with a 25nm AlGaAs current blocking layer was included 10 nm from the surface. Dots were grown using the indium flush technique, and the top dot is intentionally made thicker. A 5 nm layer of Ti was deposited prior to a 120 nm thick Al layer in which 1 μm apertures were patterned using e-beam lithography. The quantum dots are embedded in a GaAs Schottky diode which allows for deterministic charging and for tuning of energy levels through the quantum confined Stark effect. The device schematic is shown in Fig 1.

Architecture of the sample proved crucial in obtaining sample properties key to these experiments. A significant challenge was to charge two electrons in separate dots instead of the same dot. The main parameters that affect this are the relative dot sizes. The top dot was grown somewhat thicker such that it would have the lower optical transition energy and lie within the range of our lasers (~950 nm). The thickness of the bottom dot was then chosen through a series of growths to obtain the appropriate energies and charging. The nominal dot thicknesses are 2.6 and 3.2 nm for the bottom and top dot, respectively.

The second important sample parameter was the tunneling rate. Our previous set of samples with 13 nm barrier had tunneling rates of about $2t_e$=1.7 meV, resulted in singlet-triplet splittings too large relative to the optical pulse widths of our picosecond lasers. We found that larger barrier thicknesses led to a very high sensitivity to bias voltage and to voltage fluctuations. In addition, it is more of a challenge to obtain tunnel-coupled dots with barriers larger than 15 nm since at that thickness there is a lower probability of vertical stacking. Instead of the width, we chose to increase the height of the barrier by incorporating 3nm of $Al_{0.3}Ga_{0.7}As$ in between two 3 nm layers of GaAs; giving a total barrier thickness of only 9 nm. We wanted an exchange splitting $\Delta_{ee}$ of approximately 120µeV (~ 30GHz) at the bias where the 2*e* state was stable. With this value the interaction time would be very fast (of order 30ps), and we could tune our laser pulse bandwidth to both larger and smaller values. The actual anti-crossing energy is $2t_e$=700 µeV as measured from photoluminescence. With pure GaAs, a 20 nm barrier would be required.

The final crucial parameter was the spacer thickness from the bottom dot layer to the doping layer. In the experiments presented in this paper the spacer thickness is 40 nm. We also fabricated a sample with a 30 nm spacer barrier, which did not show signs of optical pumping.

**Measurement method**

The sample was patterned with an Al shadow mask with 1µm diameters apertures and placed in a He-flow cryostat at 5 K. Piezo positioners were used to position the sample in the cryostat. A 0.68 N.A. aspheric lens was used to focus the laser to a diameter of ~2 µm on the aperture, and also to collect the light for the photoluminescence measurements. Light transmitted through the aperture was focused onto an avalanche photodiode. The transitions were probed via Stark-shift modulation spectroscopy using narrow-linewidth lasers (neV range) and lock-in techniques. The sample was modulated with a 100 mV amplitude square wave voltage at 10 kHz superimposed on the fixed DC voltage. A 300ms time constant was used for all the experiments.

The single laser experiments were performed with linearly polarized light 45° to the axis of anisotropic exchange splitting. For the Rabi oscillation and Ramsey fringe experiments, one or two regenerative mode-locked Titanium:Sapphire lasers were used, with circular polarizations opposite to that of the cw probe. The pump laser(s) for all the multi-laser experiments were rejected using a polarization analyzer. For the Ramsey fringe experiments short pulses (13 ps) were detuned below the triplet transition by 270μeV, and long pulses (~150 ps), if present, were tuned resonant with the triplet. For the Rabi oscillation experiment the long pulses were tuned to the singlet transition.

The mode-locked lasers produced long or short pulses (~150 ps or 13 ps), depending on the Gires-Tournois Interferometer used for group velocity dispersion compensation. For the short pulses, the temporal pulse length was measured with an autocorrelator and the spectral bandwidth (146 μeV) with a spectrometer, giving nearly transform-limited pulses. For the long pulses, only the spectral bandwidth (10-12 μeV) could be measured using a Fabry-Perot interferometer, and the estimated pulse length is based on transform-limited Gaussian pulses.

**Simulations**

The Rabi oscillations displayed in Fig. 5b were simulated with a three-level lambda system, in which $S$ and $T_0$ couple to the same trion state (one of the $m_z = \pm 1$ states) with equal dipole moments. The three-level density matrix equations are solved numerically in response to a 120 ps Gaussian pulse resonant with $S$. A recombination time of 500 ps is used (equally to both $S$ and $T_0$), and recombination during the pulse prevents the population from being driven fully back down to $S$.

Simulations of the Ramsey fringe experiment were also performed, with 12 ps pulses detuned 300 μeV from the center of the transitions. These pulses rotated the Bloch vector from *S* toward $T_0$, with little trion population remaining after the pulse. When the delay between pulses was varied, the $T_0$ population due to the pulses oscillated as observed experimentally. However, the pulses were not short enough to completely rotate the Bloch vector all the way to the $T_0$ state. We estimate that the series of two pulses used in these experiments were able to give a maximum of 65% population in the $T_0$ state. Shorter pulses will eliminate this issue.